\documentclass[preprint,12pt]{elsarticle}
\usepackage{amssymb,amsthm,amsmath,amsfonts}
\usepackage{hyperref}
\usepackage{float}
\usepackage[margin=2cm,a4paper]{geometry}
\usepackage[usenames, dvipsnames]{color}
\usepackage[T1]{fontenc}
\usepackage{graphics}
\usepackage{epsfig}
\usepackage{listings}
\usepackage{indentfirst}
\usepackage{amssymb}
\usepackage[utf8]{inputenc}
\begin{document}
\begin{frontmatter}
\title{Functional data analysis:
Application to the second and third wave of COVID-19 pandemic in Poland}

\author[inst1]{Patrycja H\k{e}{\'c}ka \corref{cor1}}
\ead{patrycja.hecka@gmail.com}
\cortext[cor1]{Corresponding author}

\affiliation[inst1]{organization={Department of Telecommunications and Teleinformatics, Wroc\l aw University
of Science and Technology},
            addressline={Wybrze\.ze Wyspia\'nskiego 27}, 
            city={Wroc\l aw},
            postcode={50-370},
            country={Poland}}

\begin{abstract}
In this article we use the methods of functional data analysis to analyze the number of positive tests, deaths, convalescents, hospitalized and intensive care people during second and third wave of the COVID-19 pandemic in Poland. For this purpose firstly we convert the data to smooth functions. Then we use principal component analysis and multiple function-on-function linear regression model to analyze waves of COVID-19 pandemic in Polish voivodeships. 
\end{abstract}


\end{frontmatter}

\section{Introduction}\label{section:intro}

 The SARS-CoV-2 virus has become a global problem since it was revealed at the end of 2019 in China. On March 4 the first case was detected in Poland. In this paper, we use the methods of functional data analysis to analyze the number of hospitalized and intensive care people during the second and third wave of the COVID-19 pandemic in selected, Polish voivodeships. 

Determining the boundaries of individual coronavirus waves in Poland is contractual. The first case of infection was found on March 4, 2020 in a 66-year-old man in hospital in Zielona G{\'o}ra. The beginning of the first wave in Poland is therefore assumed to be spring 2020. The second COVID-19 wave in Poland lasted five months - from September to January. The peak of the second wave
fell in November with a record increase - 27,875 infections - November 7, 2020. The third wave began three months after the peak of the second wave. It is considered that the beginning of the third wave is February 16, 2021. Her record was on April 1 with 35,251 new cases of SARS-CoV-2 - the highest daily number of infections since the beginning of the pandemic in Poland. However, due to deficiencies
in the published data to the public, we assume as the second wave the period between October 23, 2020 and February 15, 2021, and as the third wave the time from February 16 to July 5, 2021. The data we used has been collected and published by Micha{\l} Rogalski (\cite{zbior}, contact:
contact.micalrg@gmail.com, see data source: \url{http://bit.ly/covid19-poland}, accessed: March 1, 2022).

\begin{figure}
\centering
\includegraphics[scale=0.7,width=0.45\textwidth]{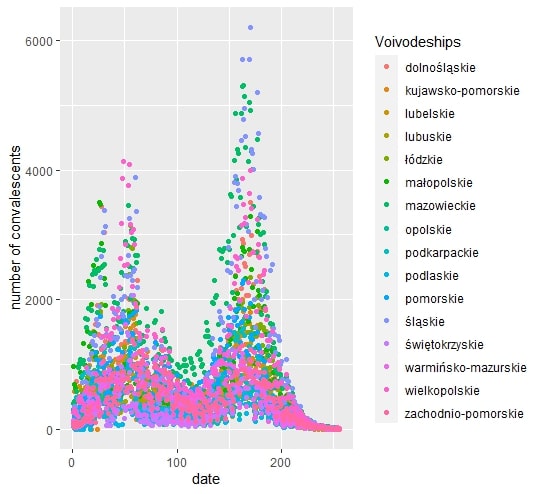}
\includegraphics[scale=0.7,width=0.45\textwidth]{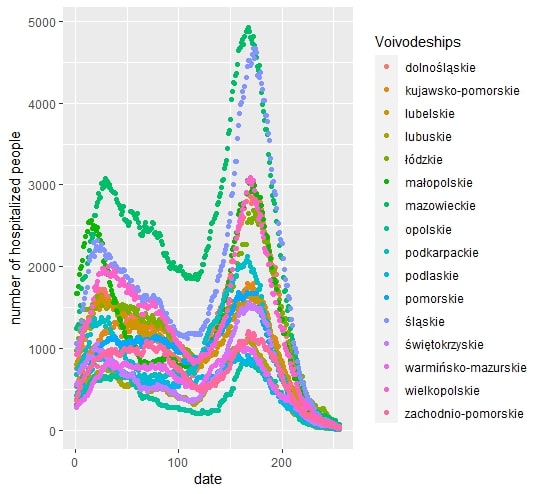}
\includegraphics[scale=0.7,width=0.45\textwidth]{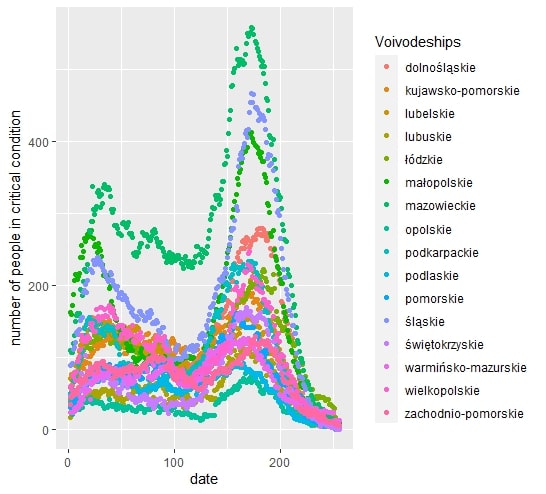}
\caption{Daily observations of the number of positive tests results, deaths, convalescents, hospitalized people and people in critical condition since October 23, 2020 to July 5, 2021 in all voivodeships in Poland.}
\label{fig:dzienneniedzielone}
\end{figure}

Figure (\ref{fig:dzienneniedzielone}) shows daily, discrete observations of the number of positive
COVID-19 test results, number of deaths, convalescents, hospitalized people and people in
serious condition.

\begin{figure}
\centering
\includegraphics[ scale=0.7,width=0.45\textwidth]{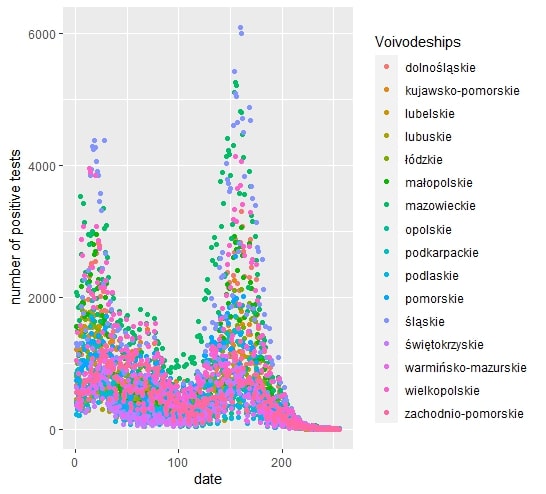}
\includegraphics[scale=0.7, width=0.45\textwidth]{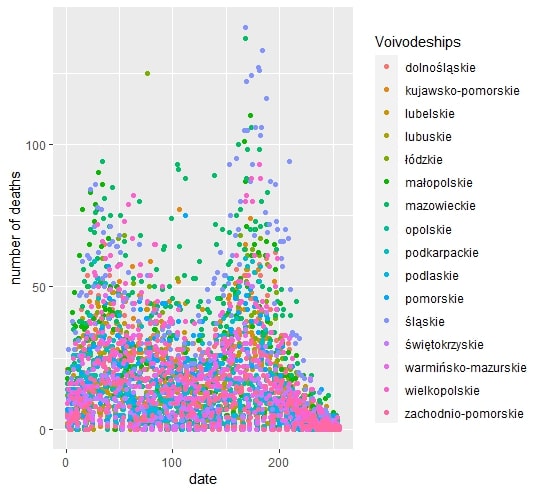}
\end{figure}

In order to reduce the influence of the number of inhabitants of a given voivodeship on the analysis the number of cases, hospitalized people, convalescents, deaths, and
people in a serious condition we divide by the number of inhabitants of a given voivodeship. Number of inhabitants we download from the website of the Statistical Information Centre (see: \cite{urzad}, \textit{''Area and population in the territorial profile in 2021, Area, population number and density, as of 1 January 2021''}, date of publication: July 22, 2021, accessed: March 7, 2022). Then the obtained numbers were multiplied by 100,000.

From the analysis of appropriately scaled values, it turns out that the largest number of hospitalized
and in serious condition people per 100,000 inhabitants between October 23, 2020 and July 5, 2021 was recorded in the \'Swi\k{e}tokrzyskie Voivodeship. Most positive tests and
deaths were noted in the Kujawsko-Pomorskie Voivodeship. The largest number of convalescents
was recorded in the Warmi\'nsko-Mazurskie Voivodeship. The least people were hospitalized
in the Wielkopolskie Voivodeship, people in a serious condition in the Pomorskie Voivodeship, deaths in the Ma{\l}opolskie Voivodeship,
positive tests in Podkarpackie, and convalescents in Podlaskie. For this reason for the test sample of the model which will be analyzed in the next chapters we select \'Swi\k{e}tokrzyskie, Wielkopolskie, Podkarpackie and Ma{\l}opolskie voivodeships.

Since then, the variability over time of positive test results, deaths, recoveries, hospitalized people and people in a serious condition will be considered through functional variables respectively: $X_1(t), X_2(t), X_3(t), Y_1(t), Y_2(t)$. Variables $X$ will be treated as predictors,
and $Y$ as response variables in functional regression models. The observed data is the number of daily values of these five functional variables for sixteen
voivodeships in Poland.

\begin{figure}
\centering
\includegraphics[ scale=0.7,width=0.45\textwidth]{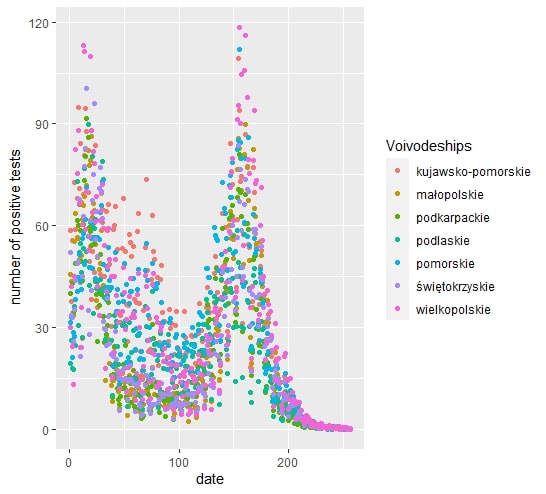}
\includegraphics[scale=0.7, width=0.45\textwidth]{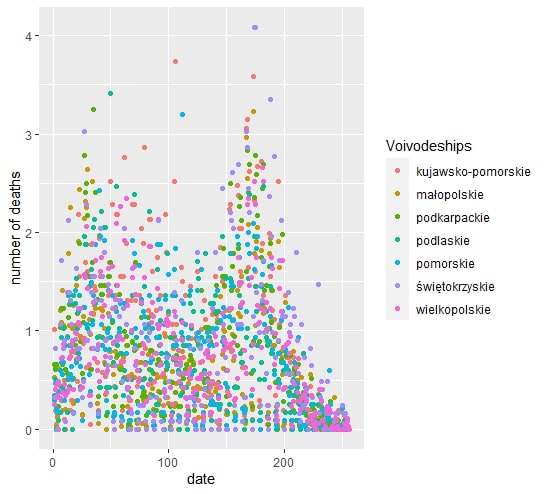}
\includegraphics[scale=0.7,width=0.45\textwidth]{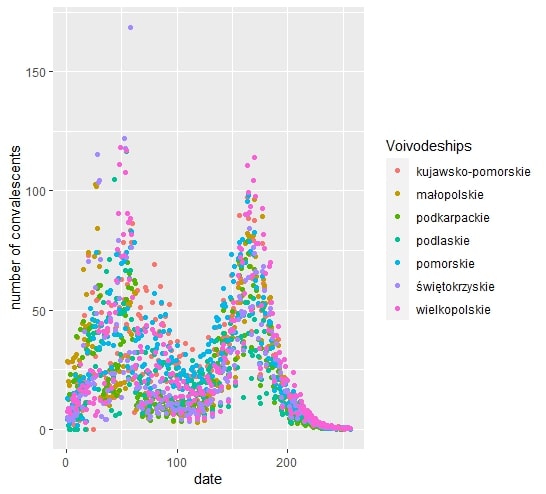}
\includegraphics[scale=0.7,width=0.45\textwidth]{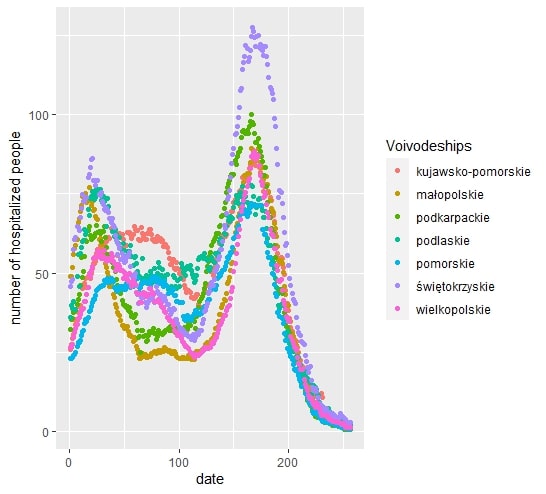}
\includegraphics[scale=0.7,width=0.45\textwidth]{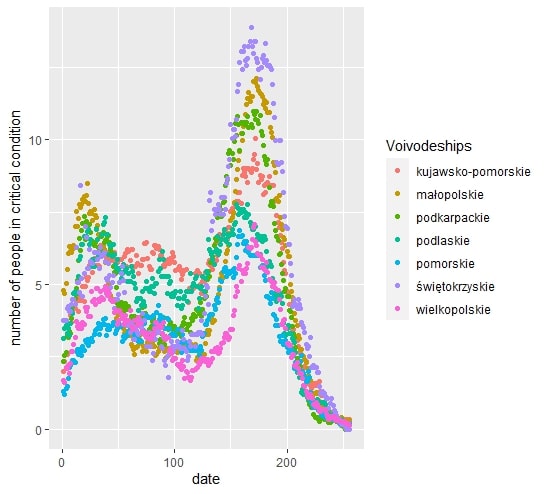}
\caption{Daily observations of the number of positive tests results, deaths, convalescents, hospitalized people and people in critical condition since October 23, 2020 to July 5, 2021 in selected voivodeships in Poland. The number of people was divided by the number of inhabitants of a given voivodeship and then
multiplied by 100,000.}
\label{fig:dziennedzielone}
\end{figure}

Figure (\ref{fig:dziennedzielone}) shows development of the disease in selected, most diverse voivodeships by discrete, daily, scaled observations,
i.e. a set of curves $\{(x_{ij} (t), y_{ik}(t)) : i = 1, ..., 16; j = 1, 2, 3; k = 1, 2\}$.

An important limitation on the FDA methods is that the functional variables
should be observed in the same interval. The classic solution to such a problem
is to transfer all curves to the same interval. In this work, we will consider
curves on the interval $T=[0,1]$. Hence, from this points $x_{ij}$ and $y_{ik}$ present the curves located at
the interval $[0,1]$.

\section{From functional data to smooth functions}\label{section:smooth_data}

There are a lot of methods of conversion data to smooth functions (see \cite{FDA_springer}, \cite{Ullah}, \cite{Boor}). Our goal is to use discrete data $y_k$, $k = 1, ..., m$ to estimate the function $x$. The main procedure in statistics, mathematics, and engineering for converting discrete data in smooth functions is the expansion of the base.

Consider the vectors  $x_i(t): i=1,...,n; t \in T=[0,1]$ and assume that the observations $y_{ik}$ are
available for each knots $t_{i1}, t_{i2}, ..., t_{im_i} \in T$. Then we can present each observation $y_{ik}$ 
as follows:
\begin{equation}
y_{ik}=x_i(t_{ik})+\epsilon_{ik}, i=1,...,n; k=1,...,m_i,    
\end{equation}
where $\epsilon_{ik}$  is a noise  contributes to the roughness of the analyzed data.

Suppose that the sample curves belong to the finite dimensional space generated by the set of basis functions $\{\phi_1(t),...,\phi_p(t)\}$.
Then we can represent each curve $x_i$ by a linear expansion of the form:

\begin{equation}
x_i(t)=\sum_{j=1}^p\alpha_{ij}\phi_j(t), i=1,...,n. \label{eq:1}
\end{equation}

We define by $\alpha_i$ vectors of the basis coefficients: $\alpha_i=(\alpha_{i1},...,\alpha_{ip})'$, which lengths are equal to {\em p} and can be estimated by different methods. One of the most popular method is least squares method (see \cite{FDA_springer}, section 4.2). The least squares estimators of $\alpha_i$ are of the form $\hat{\alpha}_i=(\Phi_i'\Phi_i)^{-1}\Phi_i'y_i,$ where $\Phi_i=(\phi_j(t_{ik}))_{m_i \times p}, j=1,...,p; k=1,...,m_i$.

Smoothness of the function  is controlled by the number of used basis functions. The greater the number of basis functions {\em p}, the better the curve fits to
discrete points. The smaller the {\em p}, the smoother the curve is. Decision
about increasing or decreasing the number of basis functions is related to achieving a compromise between bias and variance - ''bias-variance trade-off'' (see \cite{FDA_springer}, section 4.5.1). One of the methods using to get a ''trade-off'' between variance and bias is
minimalization of the mean-squared error (MSE). 

\begin{figure}
\centering
\includegraphics[ scale=0.7,width=0.45\textwidth]{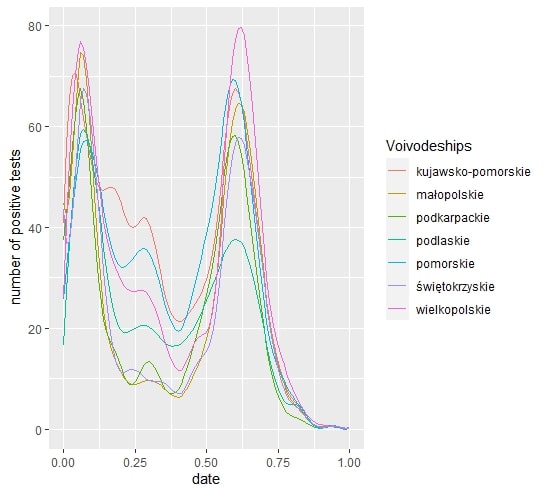}
\includegraphics[scale=0.7, width=0.45\textwidth]{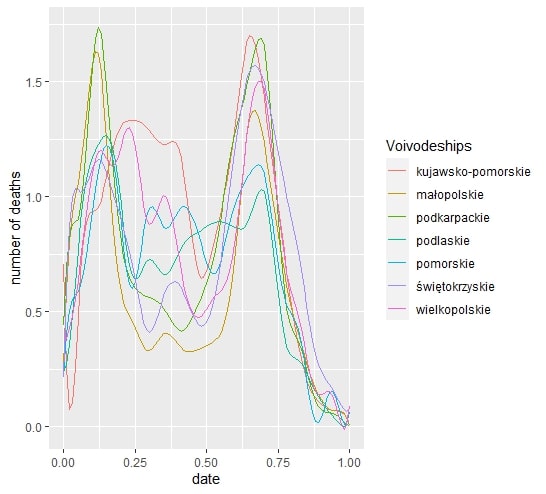}
\includegraphics[scale=0.7,width=0.45\textwidth]{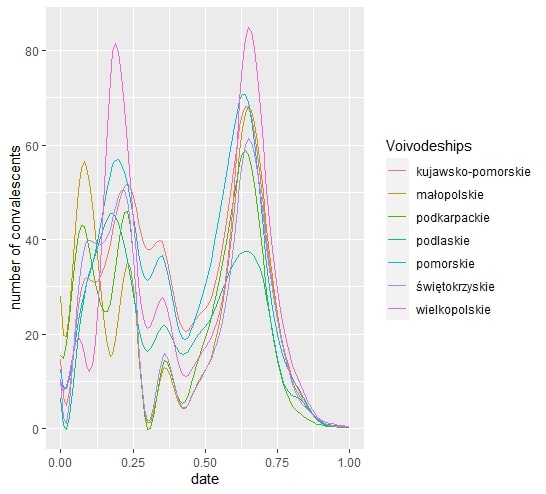}
\includegraphics[scale=0.7,width=0.45\textwidth]{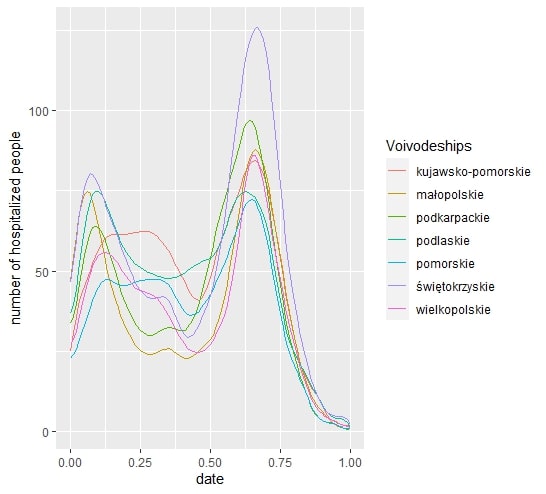}
\includegraphics[scale=0.7,width=0.45\textwidth]{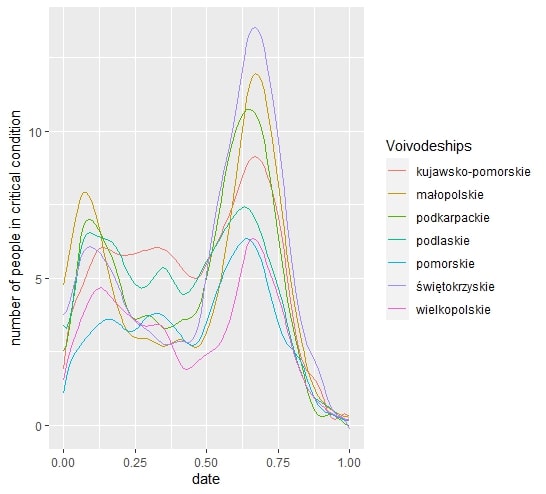}
\caption{Fitted curves to daily observations of the number of positive tests results, deaths, convalescents, hospitalized people and people in critical condition since October 23, 2020 to July 5, 2021 in selected voivodeships in Poland. 20 basis functions were used.}
\label{fig:dopasowanie}
\end{figure}

Figure (\ref{fig:dopasowanie}) shows the fitting of 20 basis functions with equally spaced knots in order to approximate seven curves showing the number of cases, deaths, recoveries, hospitalized people and people in a serious condition because of COVID-19. Coefficients for each functional form were fitted by least squares.
The number of basis functions was chosen so that the value of the mean squared error was the
smallest, but also to avoid overfitting of the model.

\section{Multiple Function-on-Function Linear Model}\label{section:MFFLR}

In this section we show multiple function-on-function linear model which was described in details e.g. by Acal, Escabias, Aguilera and Valderrama (see \cite{imputation}, section 3.1) and by Xiong Cai, Liugen Xue and Jiguo Cao  \cite{MFFLR}. Such a model is often used to characterize pandemic in Europe for example in France \cite{France}, Italy \cite{Italy} and Spain \cite{imputation}.

The MFFLR model let for estimation the functional response variable $Y$ from a vector of $J$ functional predictor variables denoted by $X=(X_1,...,X_J)'.$
Consider a random sample from  $(X,Y)$ denoted by {$(x_i,y_i):i=1,...,n$} with $x_i=(x_{i1},x_{i2},...,x_{iJ})'$. 

Then we define the functional linear model as follows:

\begin{equation}
y_i(t)=\alpha(t) + \sum_{j=1}^J \int_T x_{ij}(s)\beta_j(s,t)ds+\epsilon_i(t), i=1,...,n,  
\label{eq:MFFLR}
\end{equation}
where $\alpha(t)$ is the intercept function, $\beta_j(s,t)$ are coefficient functions and $\epsilon_i(t)$ are independent functional errors.

Consider the decomposition of the principal components of the functional response variable and the functional predictior variables, given by

\begin{equation}
x_{ij}(t)=\overline{x}_j(t)+\sum_{l=1}^{n-1} \xi_{il}^{x_j} f_l^{x_j}(t),
\label{eq:xij}
\end{equation}
\begin{equation}
y_i(t)=\overline{y}(t)+\sum_{l=1}^{n-1}\xi_{il}^yf_l^y(t),
\end{equation}
where $\xi_{il}^{x_j}$ and $\xi_{il}^y$ are the principal components scores. The eigenfunctions of the sample covariance $x_{ij}(t)$ and $y_i(t)$ are denoted by $f_l^{x_j}$ and $f_l^y$. 

The principal components decomposition given by the formula (\ref{eq:xij}) allows for the transformation of the MFFLR model described by the formula (\ref{eq:MFFLR}) into a linear regression model for each principal component
response variable $Y$ from the components of the principal functional predictors, given by the formula (\ref{eq:modelY}).

\begin{equation}
\hat{\xi}_{ik}^y=\sum_{j=1}^J \sum_{l=1}^{n-1} b_{kl}^{x_j} \xi_{il}^{x_j} + \epsilon_{ik}, i=1,...,n; k=1,...,n-1.    
\label{eq:modelY}
\end{equation}
The functional coefficients are given here by
$\beta_j(s,t)=\sum_{k=1}^{n-1}\sum_{l=1}^{n-1} b_{kl}^{x_j}f_k^{x_j}(s)f_l^y(t).$

\noindent
Finally we get the following PC-MFFLR model for the functional response:

\begin{equation}
\hat{y}_i(s)=\overline{y}(s) + \sum_{k=1}^K \hat{\xi}_{ik}^y f_k^y(s) = \overline{y}(s) + \sum_{k=1}^K(\sum_{j=1}^J \sum_{l \in L_{kj}} \hat{b}_{kl}^{x_j} \xi_{il}^{x_j})f_k^y(s),
\label{eq:PCMFFLR}
\end{equation}
where {\em K} is the number of principal components selected to the model, and $\hat{b}_{kl}^{x_j}$ are linear least-squares estimators of the regression coefficients $b_{kl}.$

Let's assume that we have {\em n} completely observed curves for all variables and {\em m} missing curves for the response variable. For missing response curves, the parameters $b_{kl}$ in model ({\ref{eq:modelY}) are estimated using complete {\em n} sample response curves and predictors. Then the missing response curves $y_i^{miss}(s): i=n+1,...,n+m$ are estimated by computing the principal components scores of the predictors: $\xi_{il}^{x_j }: i=n+1,...,n+m,l=1,...,n-1$ and inserting them into equation ({\ref{eq:PCMFFLR}}). Then the estimated PC-MFFLR model can be used to predict the value of the new response variable {\em Y} on the test sample.

We can solve the imputation problem by using the multiple function-on-function linear regression model for each response variable $Y_1(t)$ (hospitalized people) and $Y_2(t)$ (people in critical condition). Both functional regression models are
estimated from the full data of 12 voivodeships that determine the training sample. Then
predictions are made for the four, selected voivodeships:  Ma{\l}opolskie, Wielkopolskie, \'Swi\k{e}tokrzyskie and Podkarpackie.

\section{Data analysis}

In this section, we use Principal Component Analysis and a PC-MFFLR model in order to analyze functional data and predict missing response curves. The results were obtained with software R (packages 'fda' \cite{fdaR}, 'ggplot2' \cite{ggplot2}).

\subsection{Principal components analysis for functional data}

A lot of papers describe PCA in the functional context (for example see \cite{FDA_springer}, section 8).

Our first step is to estimate the principal functional components for each of the five functional predictors. It turns out that the first principal components explain respectively $46.93\%$, $42.27\%$, $38.78\%$, $41.17\%$, $44.30\%$ of the variability of $X_1$, $X_2$, $X_3$, $Y_1$, $Y_2$ from the training sample.
The second principal components explain: $26.59\%$, $19.30\%$, $26.79\%$, $29.41\%$, $31.70\%$ of the variability. The third principal components explain much less, respectively: $13.96\%$, $14.97\%$, $18.91\%$, $15.45\%$, $12.71\%$.

\subsubsection{Weight functions}

Figure ({\ref{fig:harmonics}}) presents the weight functions (harmonics) related to the first 3 principal components. The presented weight functions are coefficients that enable the eigenvectors to be computed from the original basis. 

\begin{figure}
\centering
\includegraphics[ scale=0.7,width=0.45\textwidth]{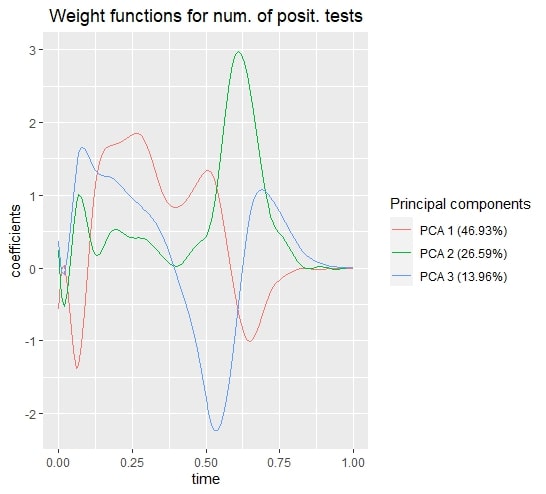}
\includegraphics[scale=0.7, width=0.45\textwidth]{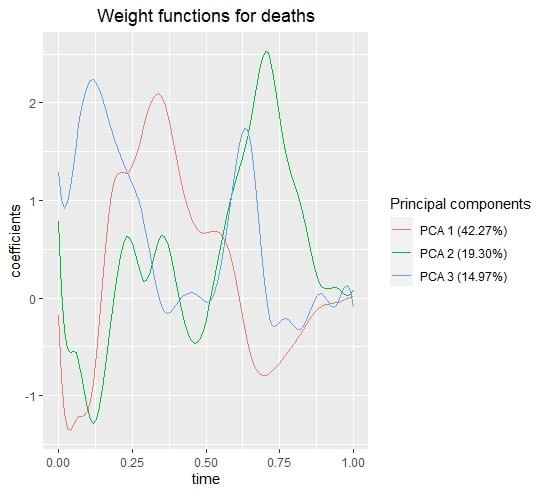}
\includegraphics[scale=0.7,width=0.45\textwidth]{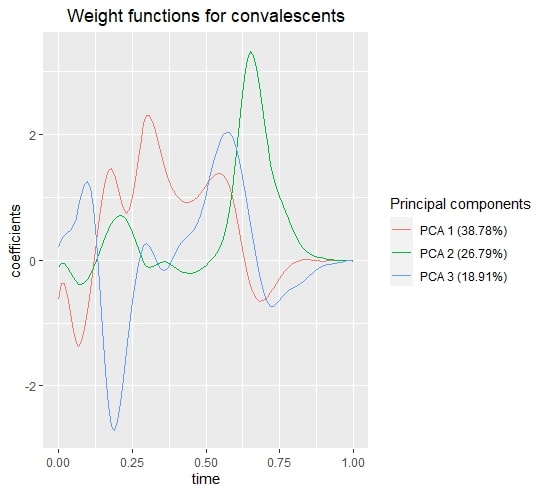}
\includegraphics[scale=0.7,width=0.45\textwidth]{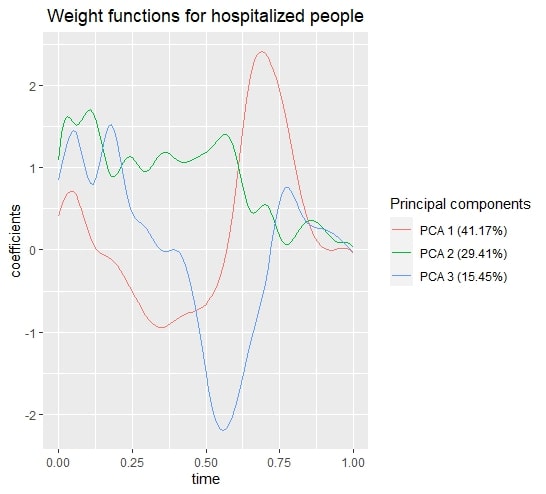}
\includegraphics[scale=0.7,width=0.45\textwidth]{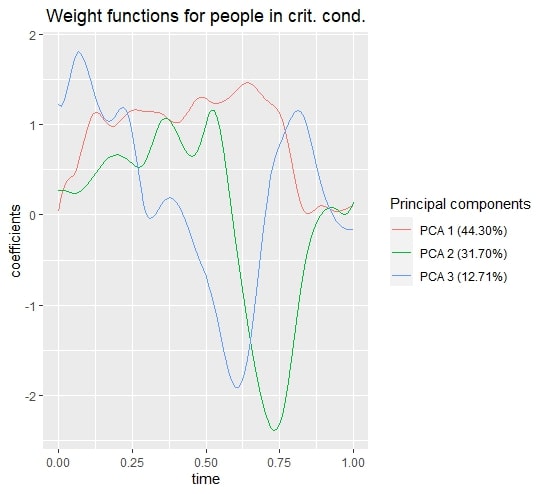}
\caption{Weight functions $f_{l}^{x_j};j=1,2,3$ and $f_l^{y_k};k=1,2$, $l=1,2,3$ for the first three principal components.}
\label{fig:harmonics}
\end{figure}

For the number of hospitalized people, the first, second and third principal components explain
together about 86.03, and for the number of people in critical condition 88.71 percentage of the variance. Other principal 
components explain a small percentage of information.
Graphs of weight functions are difficult to interpret. In the next
subsubsection we present plots that can be helpful to analyze the functional principal components. However, looking at the figure (\ref{fig:harmonics}), we can gain some intuition
about them.

The first principal component shows the general variability of the number of hospitalized people 
 depending on the wave of the pandemic. We can see that the biggest difference between the second
and the third wave in the number of hospitalized people occurs at the second half of the analyzed time -
the third wave of the COVID-19 pandemic. We can see that the smallest differences between voivodeships are
during the second wave of the pandemic. This may suggest that the prediction of the number of hospitalized people during the third wave of COVID-19 may be the hardest. In between the second
and the third wave we see negative values of the coefficients, which tell us about a decrease in the
number of  hospitalized people. Voivodeships for which the value of $\xi_{i1}^{y_1}$ is high will have
large differences between the number of hospitalized people in the second and third wave. During the third
wave, the number of hospitalized people significantly exceed the number of such people during the second wave of the pandemic. It turns out that the highest value of this coefficient is achieved for
the {\L}{\' o}dzkie Voivodeship.

Since the weight function $f_2^{y_k}; k=1,2$ must be orthogonal to $f_1^{y_k}$, we cannot expect that the second principal component will explain a greater percentage of the variance than the first
principal component. In the case of the number of hospitalized people, the second principal component explain about 29.41 percentage of variability. We see that it achieves positive values during all of the time of the second and third waves of the pandemic. The second principal component can be interpreted
as an indicator of the number of people hospitalized during the entire analyzed period.

The first principal component for the number of people in a critical condition suggests that the difference in
the number of people in a serious condition between voivodeships was at a similar level for almost the entire second and third waves of COVID-19. At this time, the values of the coefficients are positive. 
The highest value of $\xi_{i1}^{y_2}$ is obtained for the Kujawsko-Pomorskie Voivodeship.

The third and further principal components explain a much smaller proportion of variance than the first two components. This is influenced by the fact that they must be orthogonal to the first two principal components and also orthogonal to each other. They are more difficult to interpret than first two components.

Interpretation of weight functions is not always simple in the context of functional PCA. In the subsubsection (\ref{subsubsection:mean}) we show more commonly used form of presenting results.

\subsubsection{Mean curve}\label{subsubsection:mean}

A method that is helpful during the analysis of functional principal components is to present the mean curve together with the functions obtained by the addition and subtraction of the properly multiplied harmonic (weight) functions of the principal components from the mean. Such a plot makes sense because the principal components represent the variation around the mean. Figures (\ref{fig:harmonicshosp}) and (\ref{fig:harmonicsciezki})  presents the mean curves and the perturbations of the sample
mean curves obtained by adding and subtracting a multiple of weight functions for hospitalized and in critical condition people. 

\begin{figure}
\centering
\includegraphics[ scale=0.7,width=0.4\textwidth]{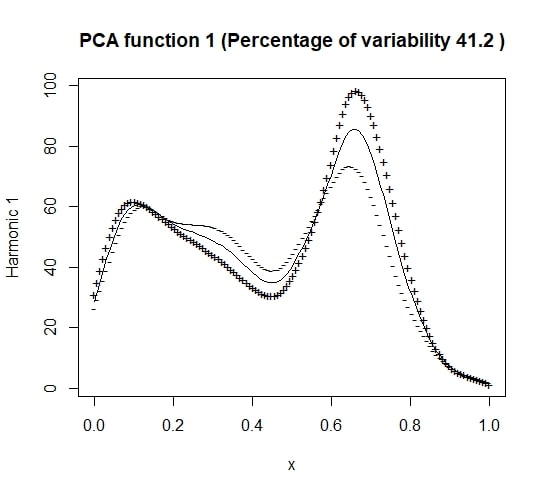}
\includegraphics[scale=0.7, width=0.4\textwidth]{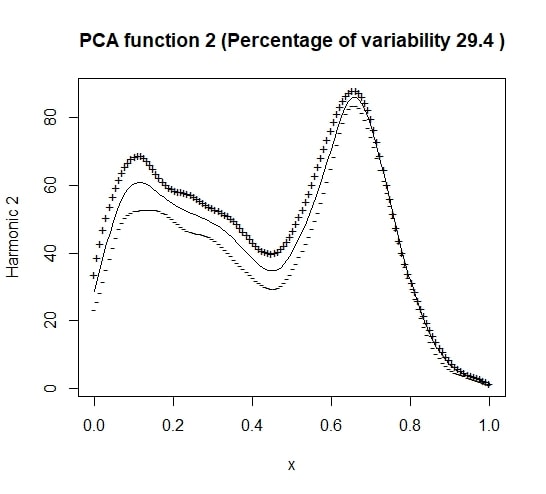}
\includegraphics[scale=0.7,width=0.4\textwidth]{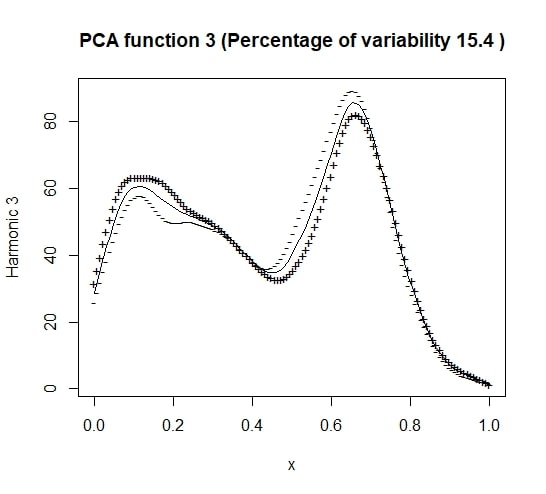}
\includegraphics[scale=0.7,width=0.4\textwidth]{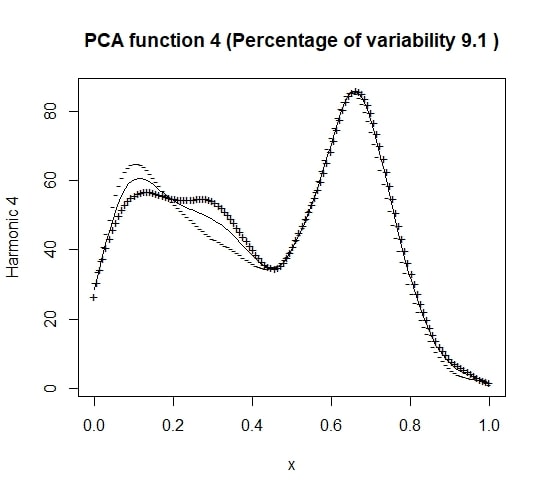}
\caption{Mean curve of hospitalized people with curves resulting from adding (+) and subtracting (-) appropriately scaled harmonic coefficients from the mean.}
\label{fig:harmonicshosp}
\end{figure}

\begin{figure}
\centering
\includegraphics[ scale=0.7,width=0.4\textwidth]{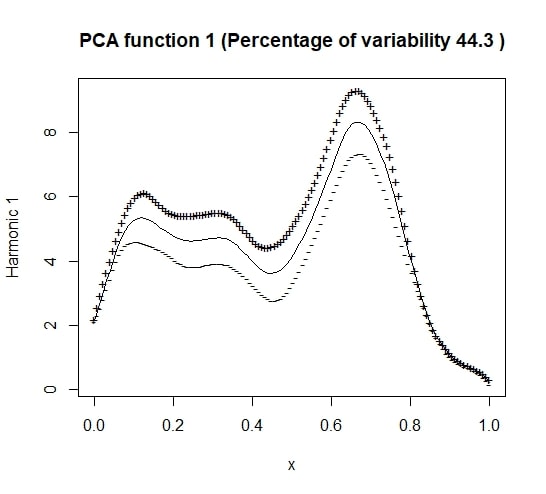}
\includegraphics[scale=0.7, width=0.4\textwidth]{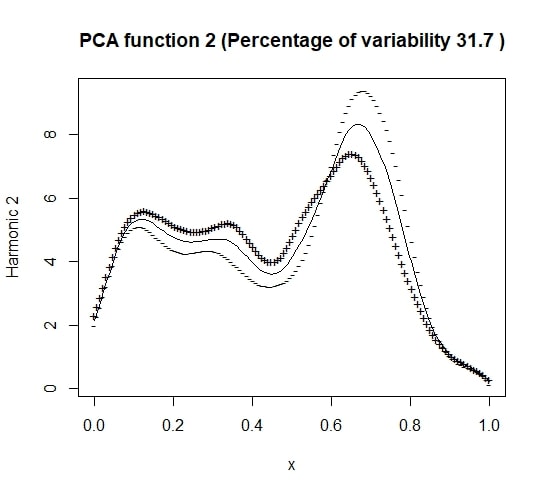}
\includegraphics[scale=0.7,width=0.4\textwidth]{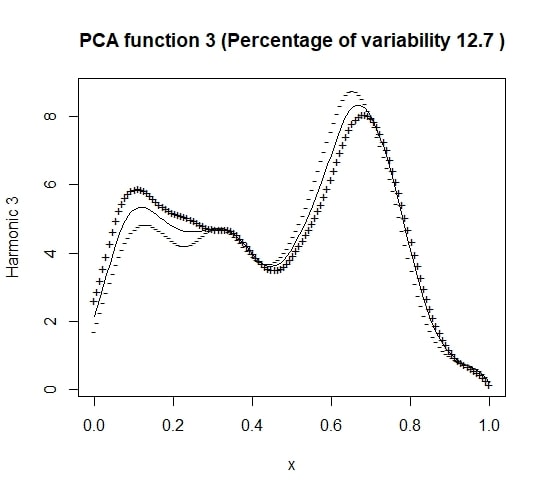}
\includegraphics[scale=0.7,width=0.4\textwidth]{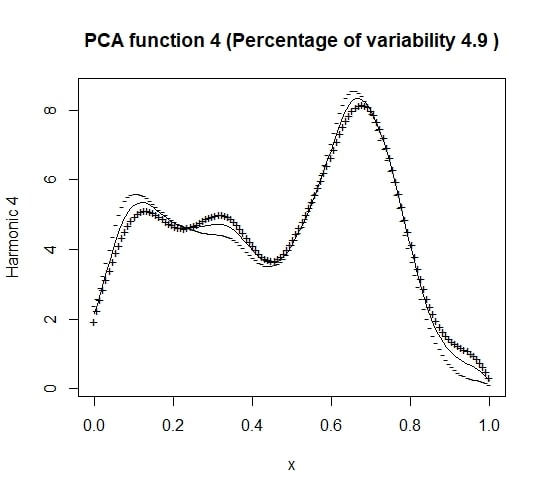}
\caption{Mean curve of people in critical condition with curves resulting from adding (+) and subtracting (-) appropriately scaled harmonic coefficients from the mean.}
\label{fig:harmonicsciezki}
\end{figure}

Analyzing the figure (\ref{fig:harmonicshosp}), we can see that in terms of the number of hospitalized people, the first principal component shows the differences between the second and third wave of the pandemic, while the second principal component shows the general number of hospitalized people during pandemic. 

The figure (\ref{fig:harmonicsciezki}) suggests that for the number of people in a serious condition the situation was reversed. The first principal component focuses on the general number of people in critical condition during pandemic, while the second principal component shows the differences between the waves. 

The next components explain a smaller percentage of variance. This can be see by examining the plots of the fourth harmonic function, where the plots of (+) and (-) often coincide with the plot of the mean. The smaller the percentage of variance is explained by the principal component, the more the (+) and (-) plots coincide with the mean function. 

\subsubsection{Plotting principal component scores}

An important aspect of PCA is the examination of the scores of each curve on each component (see {\cite{FDA_springer}, section 8.3.2).

Figures (\ref{fig:scorehosp}) and (\ref{fig:ciezkiscore}) show other graphs interesting during FPCA analysis. They represent the values of the variables achieved on the first and second scores.

\begin{figure}
\centering
\includegraphics[scale=5]{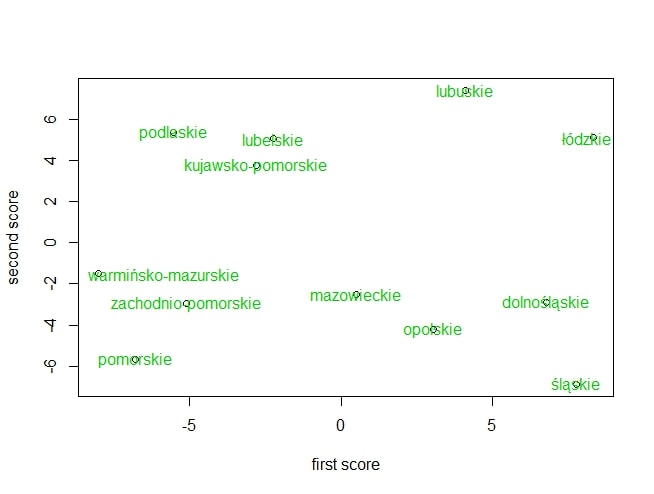}
\caption{Values of the first and second scores for the number of hospitalized people (voivodeships from the training sample).}
\label{fig:scorehosp}
\end{figure}

Suggested by the conclusions from the analysis of the figure (\ref{fig:harmonicshosp}), on the right side of the figure (\ref{fig:scorehosp}) there are voivodeships where there is the largest difference between the second and third wave of the pandemic. On the left side of the chart, those for which the difference in the number of hospitalized people is not that big in both waves. 
At the top of the figure (\ref{fig:scorehosp}), there will be voivodeships where the number of hospitalized people was the highest during the second and third wave of the pandemic. Consistently, at the bottom of the chart, we observe voivodeships with a small number of hospitalized people per 100,000 inhabitants. According to the figure (\ref{fig:scorehosp}), the largest number of hospitalized people was recorded in Lubuskie, Podlaskie, Lubelskie and {\L}{\' o}dzkie voivodeships. Comparing with appropriately scaled, discrete, real observations, the largest number of hospitalized people per 100,000 inhabitants was achieved in the following voivodeships: {\L}{\'o}dzkie, Lubuskie and Lubelskie. The voivodeships with the smallest number of hospitalized people, i.e. {\'S}l{\k a}skie, Pomorskie and Opolskie, also coincide with the real data.

Hence, we can confirm the conclusion appearing during the analysis of figure (\ref{fig:harmonics}), that the first principal component shows the differences between the second and third wave, and the second principal component is related to the number of hospitalized people. To conclude in the upper right corner, we have voivodeships with a large number of hospitalized cases and with large differences between the second and third wave. Therefore, Lubuskie and {\L}{\'o}dzkie voivodeships had the highest number of hospitalized people per 100,000 inhabitants and additionally in these voivodeships both waves differed significantly from each other.

We will now perform an analogous analysis for the figure (\ref{fig:ciezkiscore}), showing the values of the first and second scores for the number of people in a critical condition. 

\begin{figure}
\centering
\includegraphics[scale=5]{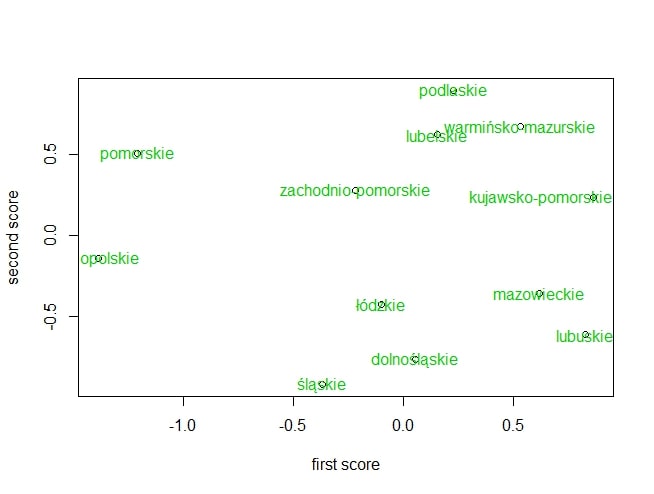}
\caption{Values of the first and second scores for the number of people in critical condition (voivodeships from the training sample).}
\label{fig:ciezkiscore}
\end{figure}

In this case, at the top of the figure (\ref{fig:ciezkiscore}), we can see the voivodeships for which the difference between the second and third wave of the pandemic is high, and low at the bottom of the figure.
On the left we have voivodeships for which the number of people in a critical condition was low, and on the right - high. Here the situation is the opposite compared to the analysis of the scores for the number of hospitalized people.

According to the data presented on the figure (\ref{fig:dziennedzielone}), the largest number of people in a serious condition per 100,000 inhabitants of a given voivodeship was reached in the Kujawsko-Pomorskie, Lubuskie and Mazowieckie voivodeships, and the smallest in: Opolskie, Pomorskie and {\'S}l{\k a}skie. Thus, we can see that the analysis of figure (\ref{fig:ciezkiscore})  seems to be performed correctly. 

Hence, the Kujawsko-Pomorskie and Lubuskie voivodeships had the largest number of people in a critical condition, while the Podlaskie and Warmi{\'n}sko-Mazurskie voivodeships had the largest differences between the second and third wave of the pandemic in the number of people in a serious condition.
Similar analyzes can be made for the number of positive test results, recoveries and deaths.

\subsection{Function-on-Function Model}

Let us consider a training sample composed of all voivodeships except \'Swi\k{e}tokrzyskie, Ma{\l}opolskie, Wielkopolskie and Podkarpackie. The listed voivodeships will be a test sample and we will make predictions for them.

Equation (\ref{eq:princcomp}) presents the reduction of the linear function-on-function to a linear model for the first principal components in terms of the first principal components of each predictor:

\begin{equation}
\hat{\xi}^{y_k}_{i1}=\gamma_0+\xi_{i1}^{x_1}\gamma_1^{y_k}+\xi_{i1}^{x_2}\gamma_2^{y_k}+\xi_{i1}^{x_3}\gamma_3^{y_k}+\epsilon_i^{y_k}, \textrm{     } k=1,2,i=1,...,16,    
\label{eq:princcomp}
\end{equation}
where $\gamma_0, \gamma_1^{y_k}, \gamma_2^{y_k}, \gamma_3^{y_k}$ are the appropriate coefficients obtained by fitting the linear model to the data.

Based on such models, we estimate the first principal components $Y_1(t)$ and $Y_2(t)$ from the first principal components of the $X_1(t)$, $X_2(t)$ and $X_3(t)$.

We predict $Y_1(t)$ and $Y_2(t)$ using the following equation (\ref{eq:predY}):
\begin{equation}
\hat{y}_{ik}(t)=\overline{y}_k(t)+\hat{\xi}_{i1}^{y_k}f_1^{y_k}(t), \textrm{    }k=1,2,i=1,...,16.   
\label{eq:predY}
\end{equation}
In order to test the models on a training sample, we will use the mean squared error.

Figure (\ref{fig:trening}) shows the graphs of the observed curves, fitted in the section (\ref{section:smooth_data}) together with the predicted curves obtained by applying formula (\ref{eq:predY}) for
several voivodeships selected from the training sample. In table (\ref{table:MSE}) you can see the values of the
mean squared error calculated for all voivodeships from the training sample.

\begin{figure}[H]
\centering
\includegraphics[scale=3]{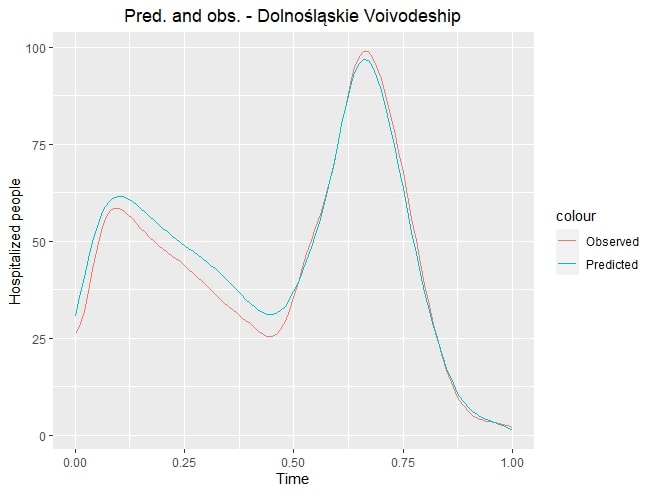}
\includegraphics[scale=3]{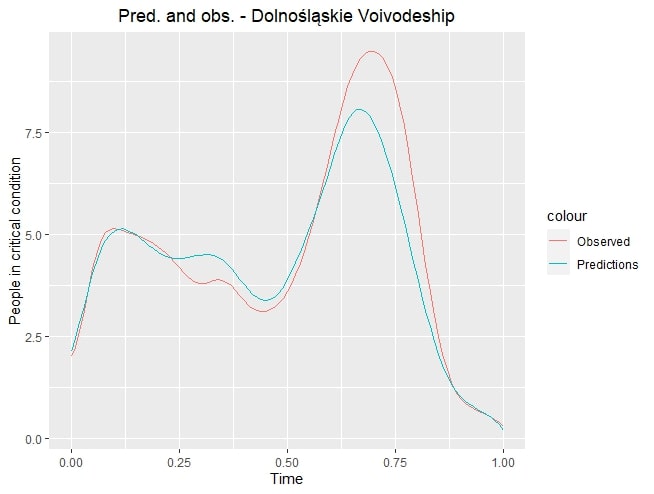}
\includegraphics[scale=3]{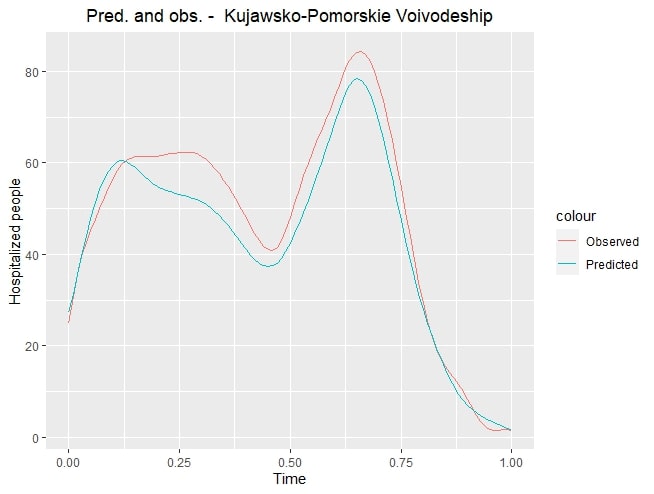}
\includegraphics[scale=3]{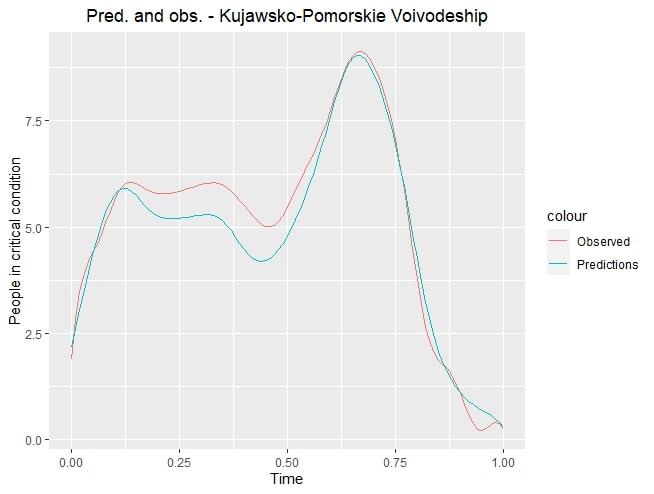}
\end{figure}
\newpage
\begin{figure}[H]
\centering
\includegraphics[scale=3]{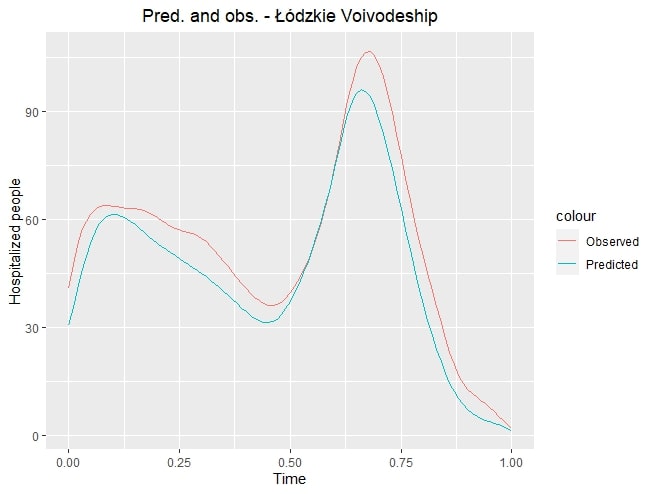}
\includegraphics[scale=3]{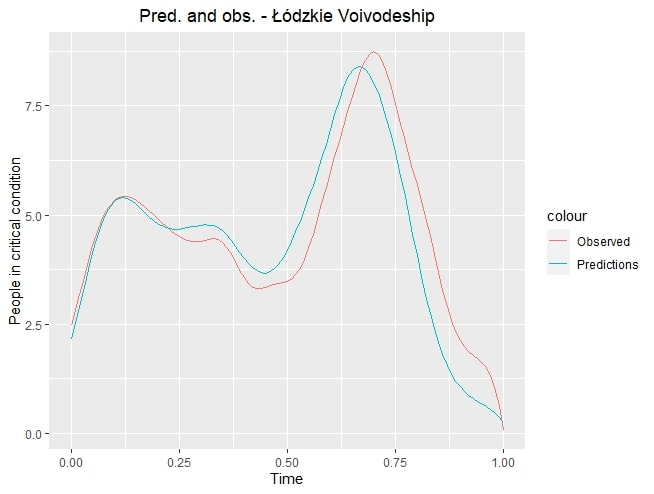}
\includegraphics[scale=3]{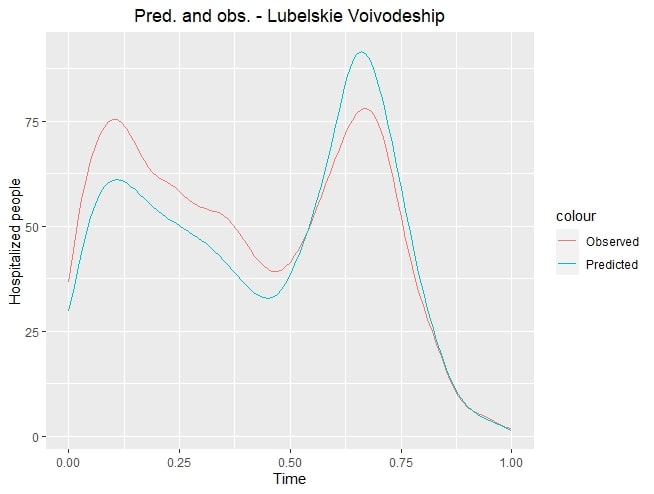}
\includegraphics[scale=3]{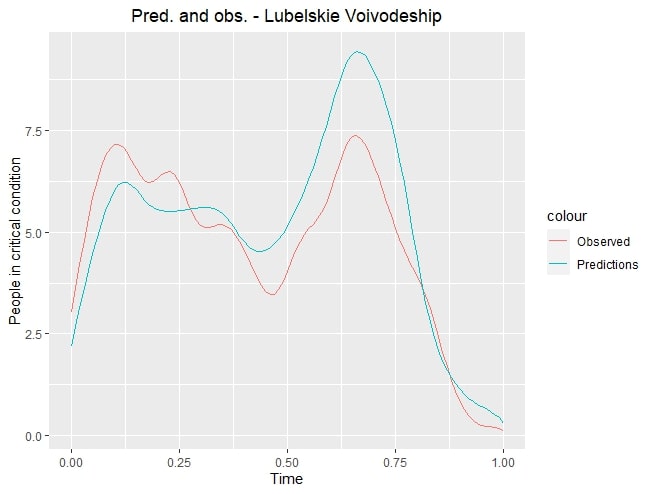}
\includegraphics[scale=3]{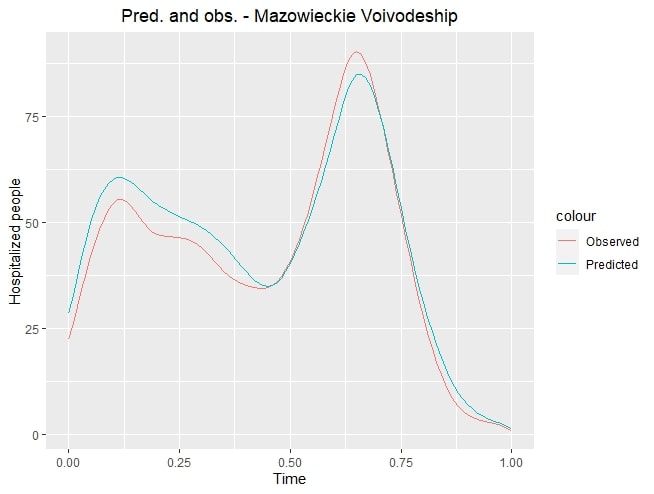}
\includegraphics[scale=3]{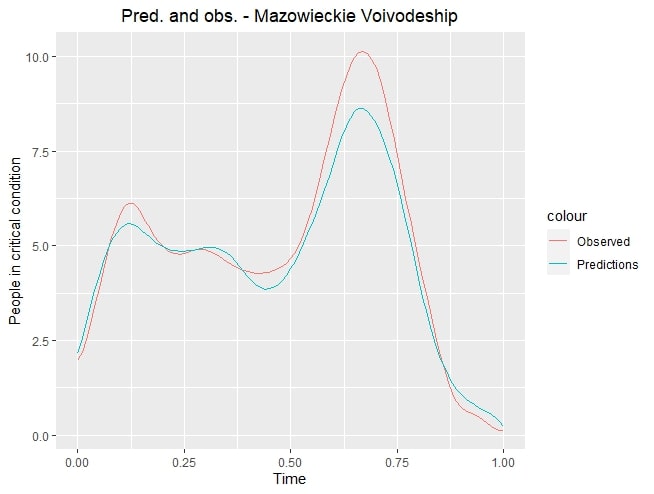}
\includegraphics[scale=3]{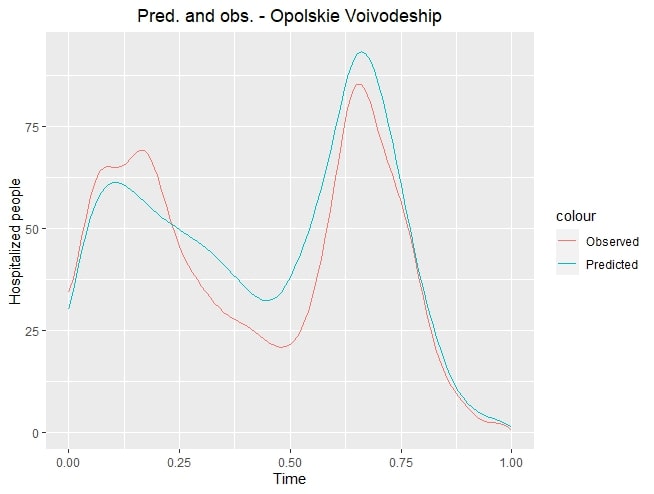}
\includegraphics[scale=3]{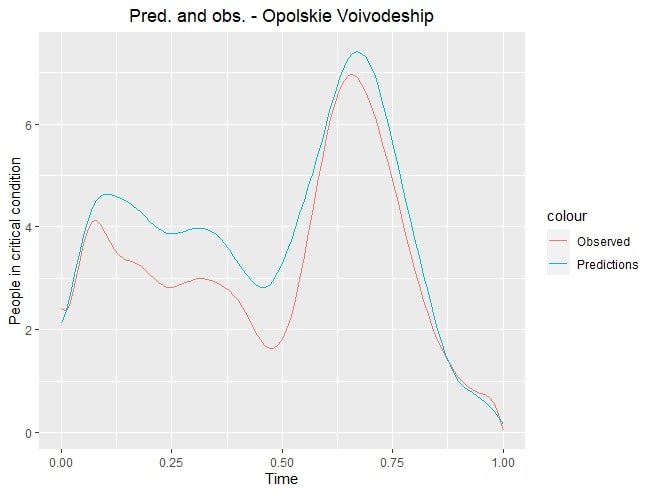}

\end{figure}
\newpage
\begin{figure}[H]
\centering
\includegraphics[scale=3]{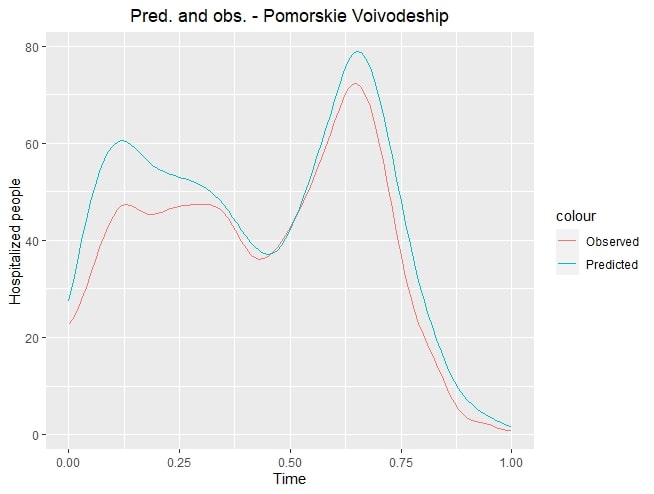}
\includegraphics[scale=3]{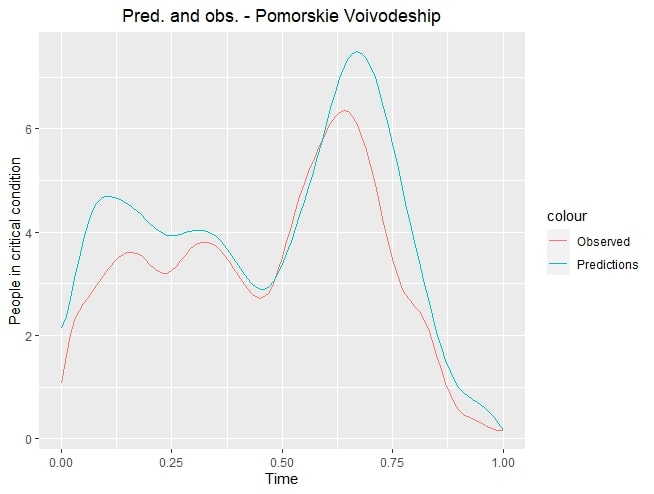}
\includegraphics[scale=3]{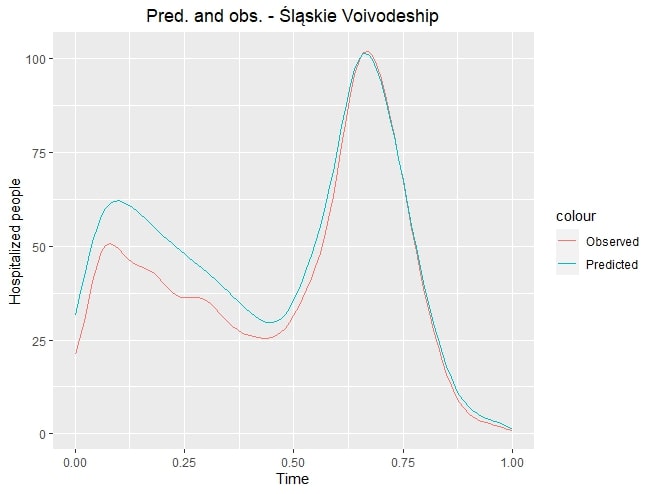}
\includegraphics[scale=3]{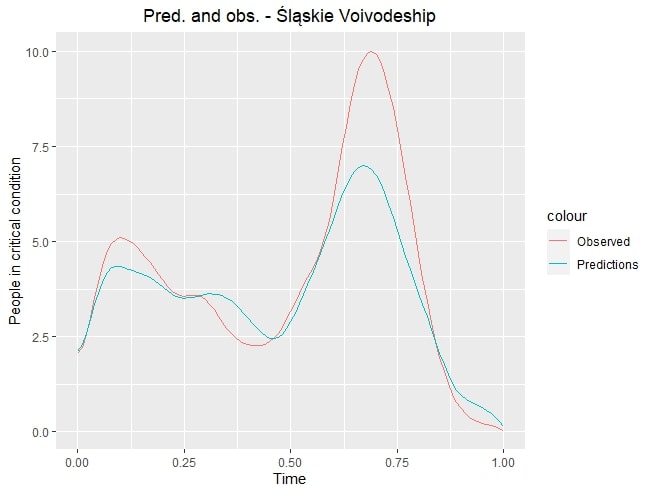}
\includegraphics[scale=3]{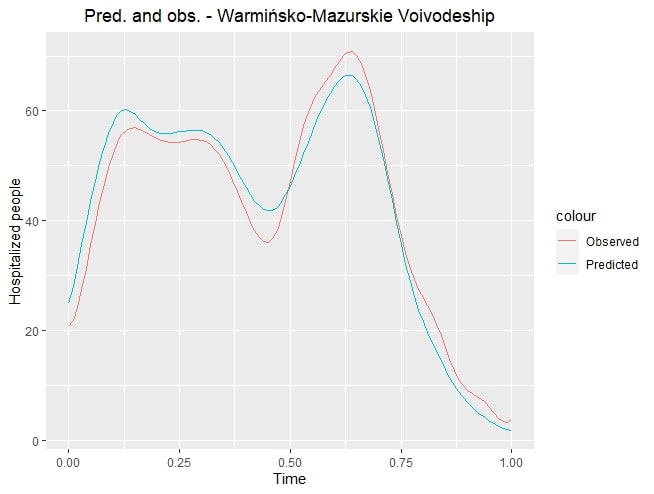}
\includegraphics[scale=3]{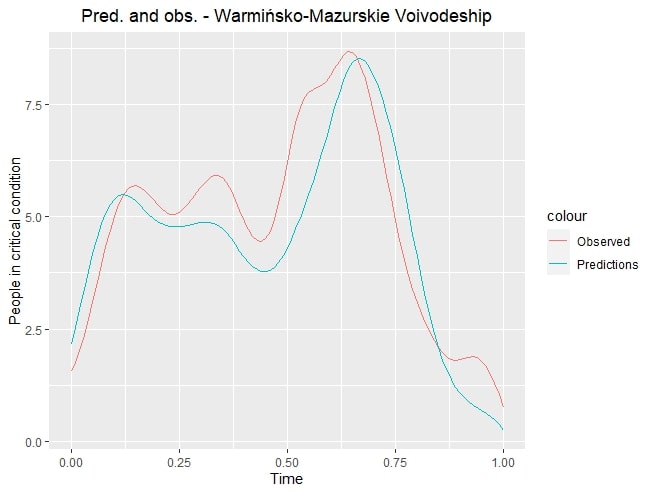}
\caption{Observed and predicted curves for selected voivodeships from the training sample.}
\label{fig:trening}
\end{figure}

\begin{table}[H]
\caption{Values of the mean squared error for $y_{i1}$ (the number of hospitalized people) and $y_{i2}$ (the number of people in a serious condition) for the voivodeships from the training sample.}
\label{table:MSE}
\centering
\begin{tabular}{ |p{6cm}||p{3cm}|p{3cm}|  }
 \hline
  voivodeship & MSE($y_{i1})$ & MSE($y_{i2}$)\\
 \hline
 dolno{\'s}l{\k a}skie   & 4.087584    &0.8424024\\
 kujawsko-pomorskie&  5.801758  & \textbf{0.5078765}\\
 {\l}{\'o}dzkie& 8.146502 & 0.7999353\\
 lubelskie & 8.434905 & \textbf{1.1796471}\\
 lubuskie & \textbf{10.79992}& 0.9807714\\
 mazowieckie & 4.569325 & 0.6134904\\
 opolskie & 9.170555  & 0.7959462\\
 podlaskie & 9.612532  & 1.0092404\\
 pomorskie & 7.526031  & 0.9541513\\
 {\'s}l{\k a}skie & 7.451072 & 1.1456427 \\
 warmi{\'n}sko-mazurskie& \textbf{3.906237} & 1.0330436\\
 zachodnio-pomorskie&  6.692964 & 0.9105502 \\
 \hline
\end{tabular}
\end{table}

In table (\ref{table:MSE}) the lowest and highest MSE values are marked in bold, respectively. We can see that the values of the mean squared error for the number of people in a serious condition are much lower than the values of the mean squared error for the number of hospitalized people. The lowest value of $\textrm{MSE}(y_{i2})$ was obtained for the Kujawsko-Pomorskie Voivodeship, and the highest for the Lubelskie Voivodeship. In turn, the value of the mean squared error for $y_{i1}$ is the lowest for the Warmi{\'n}sko-Mazurskie Voivodeship and the highest for the Lubuskie Voivodeship.

Figure (\ref{fig:testowa}) presents predictions for voivodeships from the test sample, and table (\ref{table:predandob}) presents predicted and observed values during the first and last days of the analyzed period of the pandemic. In order to compare the predictions with the true observed data, the predictions in table (\ref{table:predandob}) have been properly scaled and present the number of hospitalized and seriously ill people for voivodeship on a given day, and not the number of people per 100,000 inhabitants.

Analyzing the figure (\ref{fig:testowa}) and table (\ref{table:predandob}) we can see that the predictions for the Wielkopolskie Voivodeship turned out to be very close to the true values. The worst predictions we get for Ma{\l}opolskie Voivodeship. For example, on November 1, 102 people in a serious condition were recorded in the Wielkopolskie Voivodeship, and the model predicted 107 people; 217 were observed in the Ma{\l}opolskie Voivodeship, while the model predicted 126. On the same day, 95 people were found in a serious condition in the Podkarpackie Voivodeship, against the expected 76, and in the \'Swi\k{e}tokrzyskie Voivodeship, 55 against 41.
In turn, the number of hospitalized people on November 1 in the Ma{\l}opolskie Voivodeship amounted to 2,366 people, and the model predicted 1,591 cases; in the Podkarpackie Voivodeship 976 against the predicted 977; in the \'Swi\k{e}tokrzyskie Voivodeship 829 were observed against 589 predicted, and in the Wielkopolskie Voivodeship, 1,292 against 1,700.

Analyzing the figure (\ref{fig:testowa}) and table (\ref{table:predandob}) we can see that the closest results to the observed values were predicted at the beginning and at the end of the analyzed time. The largest difference between the observed and predicted values is for the Ma{\l}opolskie Voivodeship. Here the model sometimes predicts almost 2 times lower values than observed.

\begin{figure}[H]
\centering
\includegraphics[scale=3]{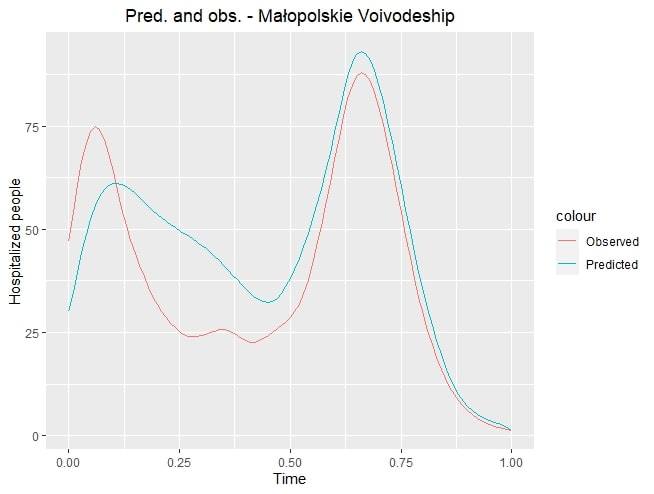}
\includegraphics[scale=3]{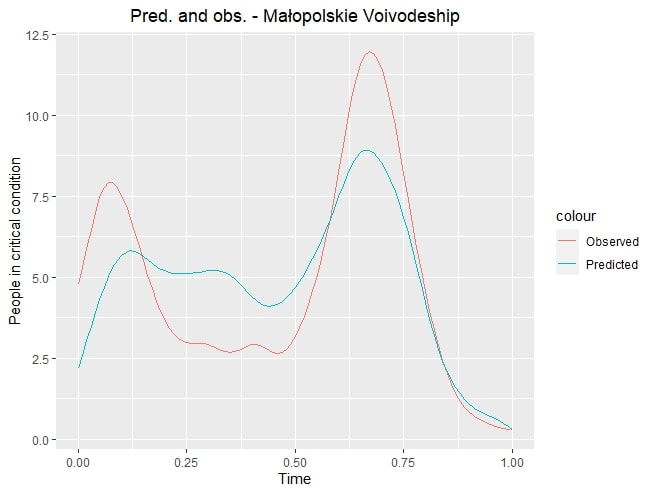}
\includegraphics[scale=3]{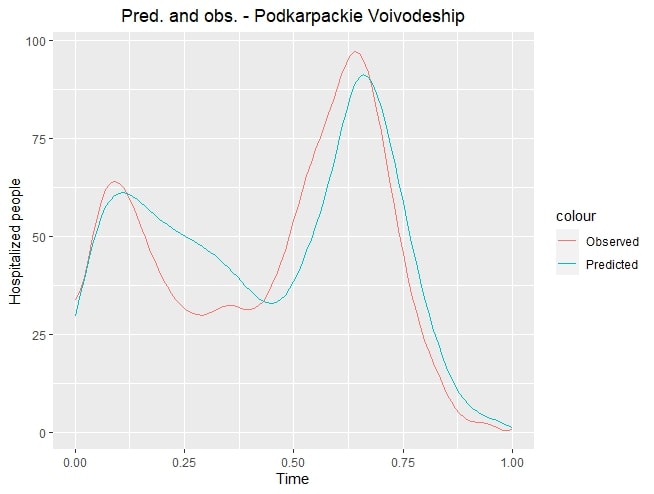}
\includegraphics[scale=3]{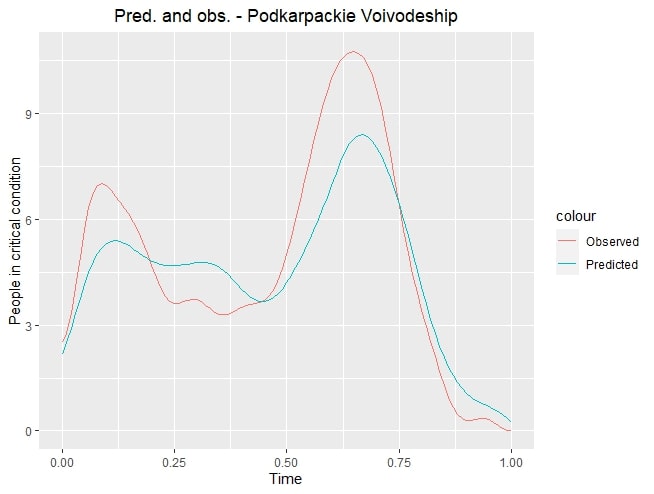}
\includegraphics[scale=3]{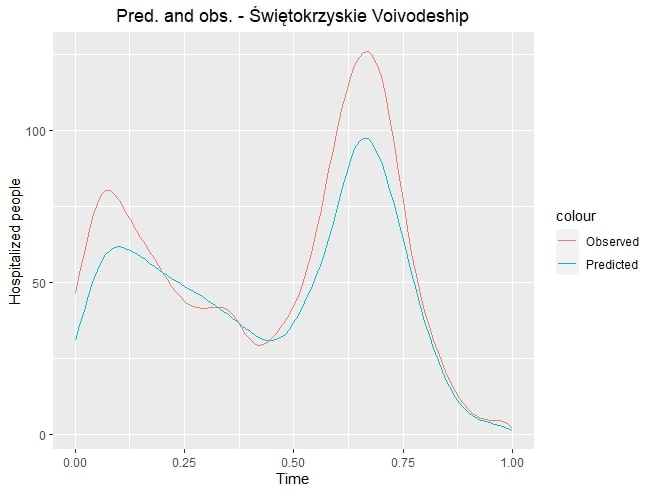}
\includegraphics[scale=3]{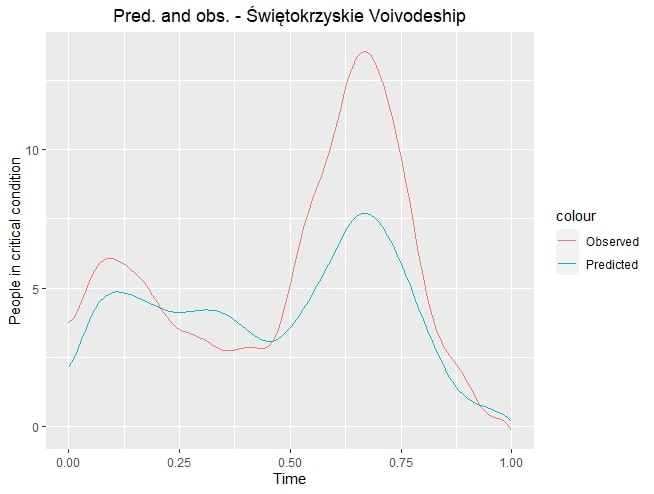}
\includegraphics[scale=3]{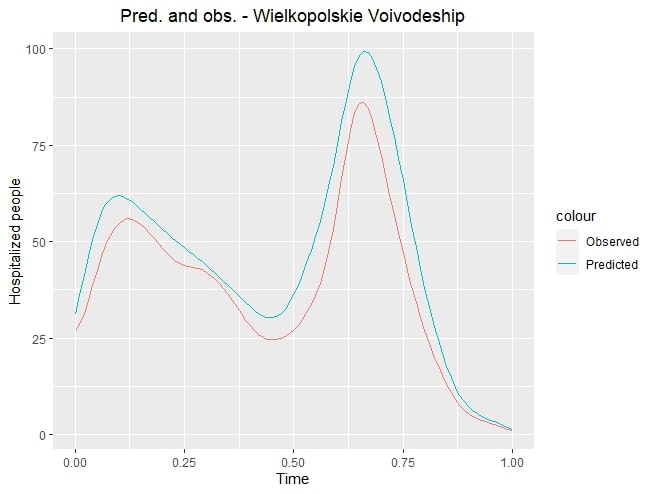}
\includegraphics[scale=3]{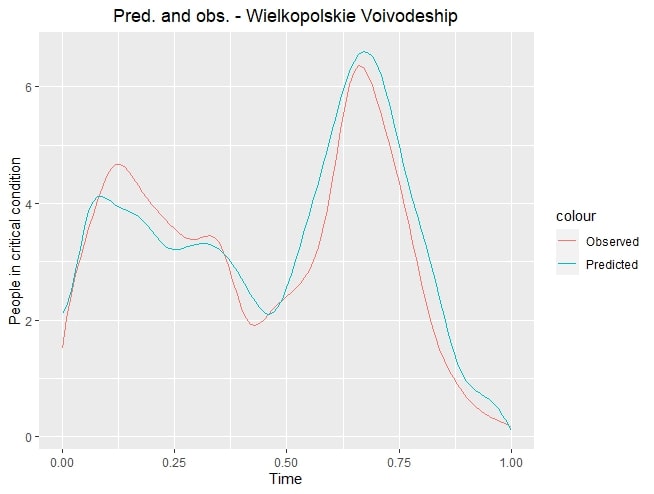}
\caption{Observed and predicted curves for the test sample.}
\label{fig:testowa}
\end{figure}

\begin{table}[H]
\caption{Predicted values for the number of hospitalized people and in the serious condition compared to the observed values. Observations are denoted by ''obs'' and predictions by ''pred''.}
\begin{tabular}{ |p{3cm}||p{3cm}|p{3cm}|p{3cm}|p{3cm}|  }
 \hline
 \multicolumn{5}{|c|}{Hospitalized people} \\
  \hline
  & ma{\l}opolskie & podkarpackie &\'swi\k{e}tokrzyskie&wielkopolskie\\
 \hline
 time & obs/pred & obs/pred &obs/pred & obs/pred\\
 \hline
 23.10  & 1671/1023    &681/629&  559/377& 924/1088\\
 24.10 &  1667/1086  & 750/668   & 581/401 &929/1158\\
 25.10 & 1763/1151 & 757/707&  589/426&961/1229\\
 26.10 & 1907/1216 & 796/747&  704/450&1039/1300\\
 27.10 & 1972/1281  & 819/786&  742/474&1090/1370\\
 28.10 & 2003/1345  & 813/826   &750/498&1137/1439\\
 29.10 & 2072/1408  & 848/865& 792/522 &1182/1507\\
 30.10 & 2250/1471 & 883/903 & 789/545 &1241/1574\\
 31.10 & 2289/1532 & 917/941 & 779/567 & 1252/1638\\
 1.11& 2366/1591 & 976/977& 829/589 & 1292/1700\\
 ... & ... &...& ...&...\\
 26.06& 72/106& 31/66& 69/39&81/111\\
 27.06& 72/100& 26/62& 61/37&79/105\\
 28.06& 66/95& 25/58 & 57/34&82/99\\
 29.06& 63/89& 20/55& 54/32&82/93\\
 30.06& 57/82& 15/51& 51/30& 63/86\\
 1.07& 53/76& 13/47& 47/27& 42/79\\
 2.07& 51/69 & 13/43& 39/25&44/71\\
 3.07& 53/62& 13/39& 36/22 &44/62\\
 4.07& 48/54 & 15/34 & 32/19&44/53\\
 5.07& 41/46 & 14/29& 33/15& 39/43\\
 \hline
\end{tabular}
\begin{tabular}{ |p{3cm}||p{3cm}|p{3cm}|p{3cm}|p{3cm}|  }
 \hline
 \multicolumn{5}{|c|}{People in critical condition} \\
 \hline
 23.10   & 163/74    &50/46&  46/26& 59/74\\
 24.10&   160/80  & 54/49   &46/27&56/75\\
 25.10& 172/85 & 56/52&  46/29&75/77\\
 26.10& 188/91 & 68/55&  51/30&74/80\\
 27.10& 207/97  & 67/58& 52/32&77/83\\
  28.10& 215/103  & 72/62   &50/34&68/87\\
  29.10& 209/109  & 70/65& 50/35 &80/92\\
  30.10& 208/115 & 81/69 & 53/37 &94/97\\
  31.10& 212/121& 90/72 & 55/39 & 94/102\\
 1.11& 217/126& 95/76& 55/41& 102/107\\
 ... & ... &...& ...&...\\
 26.06& 11/21& 5/13& 4/7&11/18\\
 27.06& 11/20& 4/12& 3/7&10/17\\
 28.06& 14/19&3/11 & 2/6&9/16\\
 29.06& 12/18& 2/11& 2/6&9/14\\
 30.06& 11/17& 2/10& 2/5&9/13\\
 1.07& 11/16& 0/9& 1/5&7/11\\
 2.07&11/14 & 0/8& 0/4&7/9\\
 3.07& 11/13& 0/7& 0/4&7/8\\
 4.07& 11/11& 1/6 &0/3 &7/6\\
 5.07&10/9 & 0/5& 0/2&5/4\\
 \hline
\end{tabular}
\label{table:predandob}
\end{table}

\section{Conlusions}
The COVID-19 pandemic has shocked the whole world. An epidemic of this scale is relatively a new phenomenon, which is why it has attracted the attention of a large number of analysts around the world. 
In the presented work, our target was to fit a model based on functional analysis data to data of the second and third wave of the COVID-19 pandemic in Poland. The aim of the model was to predict the number of people in a serious condition and hospitalized people. The following were used as predictors: the number of deaths, convalescents and positive test results. Estimations of the parameters were made on the training group consisting twelve voivodeships. Then we test the quality of these parameters. Next, predictions were made for the voivodeships: Ma{\l}opolskie, Podkarpackie, \'Swi\k{e}tokrzyskie and Wielkopolskie. 

Model predicts numbers well in most voivodeships. The biggest problem turned out to be Ma{\l}opolskie Voivodeship. The best predictions we observed for Wielkopolskie and \'Swi\k{e}tokrzyskie voivodeships. The analysis of principal components turned out to be interesting. The figures (\ref{fig:scorehosp}) and (\ref{fig:ciezkiscore}) showed that during the second and third wave, most people were hospitalized in the Lubuskie and Podlaskie voivodeships and the largest number of people in a serious condition was in voivodeships Kujawsko-Pomorskie and Lubuskie. The biggest difference between the second and third waves in the number of people hospitalized was
in the {\L}{\'o}dzkie and {\'S}l{\k a}skie voivodeships. The biggest difference between the second and third wave in the number of people in serious condition was in Podlaskie and Warmi{\'n}sko-Mazurskie voivodeships.

\section*{Acknowledgements}
 The author is grateful to Prof. Krzysztof Topolski (University of Wroc{\l}aw) for helpful comments on the master's thesis. 

 \section*{Funding}
 This research did not receive any specific grant from funding agencies in the public, commercial, or not-for-profit sectors.

\section*{Declaration of interest}
None.

\end{document}